\documentclass[letterpaper, 11pt]{article}

\usepackage{epsfig}
\usepackage{amsmath, amsthm, amssymb}
\usepackage{graphicx}
\usepackage[dvips]{hyperref} 

\def\~{$\approx$}
\def\ie{\emph{i.e.} }
\def\r{\theta}
\def\s{r}
\def\cs{C_{swaps}}
\def\cc{C_{conn}}
\def\ct{C}
\def\mt{\mathbf{T}}

\newcommand{\ccaption}[1]{\caption{\protect\parbox[t]{0.7\textwidth}{#1}}}

\newtheorem{thm}{Theorem}
\newtheorem{cor}[thm]{Corollary}
\newtheorem{emp}{Empirical Result}
\newtheorem{hyp}{Hypothesis}
\newtheorem{df}{Definition}
\newtheorem{heu}{Heuristics}
\newtheorem{con}{Convention}
\newtheorem{conj}{Conjecture}
\newtheorem{lem}[thm]{Lemma}

\newcommand{\1}{(\textbf{i})}
\newcommand{\2}{(\textbf{ii})}

\begin{document}

\begin{center}
\Large
  Fast generation of random connected graphs
  with prescribed degrees

\vspace{1cm}

\normalsize
 Fabien Viger\footnotemark[1]$^,$\footnote{LIP6, University Pierre and Marie Curie, 4 place Jussieu, 75005 Paris},
 Matthieu Latapy\footnote{LIAFA, University Denis Diderot, 2 place Jussieu, 75005 Paris}

\{fabien,latapy\}@liafa.jussieu.fr

\end{center}

\abstract

We address here the problem of generating random graphs uniformly from the
set of simple connected graphs having a prescribed degree sequence.
Our goal is to provide an algorithm designed
for practical use
both because of its ability to generate very large graphs (efficiency) and
because it is easy to implement (simplicity).

We focus on a family of heuristics for which we prove optimality conditions,
and show how this optimality can be reached in practice.
We then propose a different approach, specifically designed for typical
real-world degree distributions, which outperforms the first one. 
Assuming a conjecture which we state and argue rigorously, we finally
obtain an $O(n \log n)$ algorithm, which, in spite of being very simple,
improves the best known complexity.

\section{Introduction}
 
In the context of large complex networks, the generation of random\footnote{
  In all the paper, \emph{random} means \emph{uniformly at random}: each
  graph in the considered class is sampled with the same probability.} 
graphs is intensively used for simulations of various kinds.
Until recently, the main model was the Erd\" os and Renyi
\cite{erdos-renyi, bollobas} one. Many recent studies however gave evidence of the fact
that most real-world networks have several properties in common
\cite{newman-review, barabasi-review, dorogovtsev, newman-strogatz}
which make them very different from random graphs.
Among those, it appeared that the degree distribution of most real-world
complex networks is well approximated by a power law, and that this
unexpected feature has a crucial impact on many phenomena of interest
\cite{clauset-moore, newman-strogatz, newman-review, faloutsos}.
Since then, many models have been introduced to capture this feature.
In particular, the Molloy and Reed model \cite{molloy-reed}, on which we
will focus, generates a random graph with prescribed degree sequence
in linear time.
However, this model produces graphs that are neither \emph{simple}\footnote{
  A simple graph has neither multiple edges, \ie several edges binding the
  same pair of vertices, nor loops, \ie edges binding a vertex to itself.}
nor \emph{connected}.
To bypass this problem, one generally simply removes multiple edges and loops,
and then keeps only the largest connected component. Apart from the expected
size of this component \cite{molloy-reed2, aiello}, very little is known about the impact
of these removals on the obtained graphs,
on their degree distribution and on the simulations processed using them.

The problem we address here is the following: given a degree sequence, we want
to generate a random \emph{simple connected} graph having exactly this
degree sequence.
Moreover, we want to be able to generate very large such graphs, typically with
more than one million vertices, as often needed in simulations.

Although it has been widely investigated, it is still an open problem to
directly generate such a random 
 graph, or even to enumerate them in polynomial time,
even without the connectivity requirement \cite{rao, roberts, milo}.

In this paper, we will first present the best solution proposed so far
\cite{gkantsidis, milo},
discussing both theoretical and practical considerations.
We will then deepen the study of this algorithm, which will lead us
to an improvement that makes it optimal among its family.
Furthermore, we will propose a new approach solving the problem in
$O(n \log n)$ time, and being very simple to implement.

\section{Context}

\subsection*{The Markov chain Monte-Carlo algorithm}
Several techniques have been proposed to solve the
problem we address. We will focus here on the Markov chain Monte-Carlo 
algorithm \cite{gkantsidis},
pointed out recently by an extensive study \cite{milo}
as the most efficient one.

\noindent
The generation process is composed of three main steps:
\begin{enumerate}
\item \textbf{Realize the sequence}:
  generate a simple graph that matches the degree sequence,
\item \textbf{Connect}
  this graph, without changing its degrees, and
\item \textbf{Shuffle}
  the edges to make it random, while keeping it connected and simple.
\end{enumerate}

The Havel-Hakimi algorithm \cite{havel, hakimi} solves the first step
in linear time and space.
A result of Erd\"os and Gallai \cite{erdos-gallai} shows that this
algorithm succeeds if and only if the degree sequence is realizable.

The second step is achieved by swapping edges to merge separated connected
components into a single connected component, following a well-known
graph theory algorithm \cite{berge,taylor}.
Its time and space complexities are also linear.

\begin{figure}[!hb]
\begin{center}
\includegraphics[height=1.4cm]{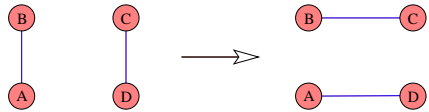}
\end{center}
\caption{Edge swap}
\label{fig:swap}
\end{figure}

The third step is achieved by randomly swapping edges of the graph,
checking at each step that we keep the graph simple and
connected. Given the graph $G_t$ at some step $t$, we pick two edges
at random, and then we swap them as shown in Figure~\ref{fig:swap},
obtaining another graph $G'$ with the same degrees.
If $G'$ is simple and connected, we consider the swap as \emph{valid}:
$G_{t+1} = G'$. Otherwise, we reject the swap: $G_{t+1} = G_t$

This algorithm is a Markov chain where the
\textbf{space} $S$ is the set of all simple connected graphs
with the given degree sequence,
the \textbf{initial state} $G_0$ is the graph obtained by the first
two steps, and
the \textbf{transition} $G_t \to G_{t+1}$ has probability $\frac{1}{m(m-1)}$
  if there exists an edge swap that transforms $G_t$ in $G_{t+1}$.
  If there are no such swap, this transition has probability $0$
  (note that if $G_t = G_{t+1}$, the probability of this transition is
  given by the number of swaps that disconnect the graph divided by $m(m-1)$).

We will use the following known results:
\begin{thm}
This Markov chain is irreducible \cite{taylor}, symmetric \cite{gkantsidis},
  and aperiodic \cite{gkantsidis}.
\end{thm}

\begin{cor} \label{cor:markov-conv}
The Markov chain converges to the uniform distribution on
  every states of its space, \ie all graphs having the wanted properties.
\end{cor}

These results show that, in order to generate a random graph, it is sufficient
to do \emph{enough} transitions.
However, no formal result is known about the convergence
speed of the Markov chain, \ie the required number of transitions.
A result from Will \cite{will}
bounds the diameter of the space $S$ by $m$.
Furthermore, massive experiments \cite{gkantsidis, milo}
showed clearly that, even if the original graph (initial state) is
extremely biased, $O(m)$ transitions are sufficient to make the graph appear to
be ``really'' random. More precisely, the distributions of a large set of
non-trivial metrics (such as the diameter, the flow, and so on) over
the sampled graphs is not different from the distributions obtained with
random graphs. Notice that we tried, unsuccessfully, to find a metric
that would prove this assertion false. Therefore, we will assume the following:

\begin{emp}\cite{milo,gkantsidis} \label{emp:conv}
The Markov chain converges after $O(m)$ swaps.
\end{emp}

\subsection*{Equivalence between swaps and transitions}
Notice that Empirical Result~\ref{emp:conv} concerns actual swaps, and
not the transitions of the Markov chain: in order to do $O(m)$ swaps one may
have to process much more transitions. This point has never been discussed
rigorously in the literature, and we will deepen it now.
We proved the following result:
\begin{thm} \label{th:swaps-transitions}
  For any simple connected graph, let us denote by $\rho$ the fraction
  of all possible pairs of vertices which have distance greater than or equal
  to $3$.
  The probability that a random edge swap is \textbf{valid}
  is at least $\frac{\rho}{2z(z+1)}$, where $z$ is the average degree.
\end{thm}

\begin{proof}
  If $\rho=0$ the result is trivial. If $\rho>0$, consider a pair $(v,w)$
  of vertices having distance $d(v,w) \geq 3$ (\ie
  there exist no path of length lower than 3 between $v$ and $w$).
  Since the graph is connected, there exists a path of length $l \geq 3$
  \ $\left(v, v_1, \cdots, v_{l-1}, w \right)$ \ connecting $v$ and $w$.
  The edge swap $(v,v_1) (w,v_{l-1}) \to (v,v_{l-1}) (w,v_1)$ is
  \textbf{valid}: it does not disconnect the graph, and since the edges it
  creates could not pre-exist (else we would have $d(v,w)\leq 2$),
  it keeps it simple.
  
  Now, the $\rho \cdot n(n-1)$ ordered pairs of vertices define at least
  $\frac{\rho \cdot n(n-1)}{8}$ edges swaps, since an edge swap corresponds
  to at most $8$ ordered pairs. Therefore, a random edge swap is valid with
  probability at least $\frac{\rho \cdot n(n-1)}{8 m(m-1)}$.
  The fact that $m = \frac{n \cdot z}{2}$ ends the proof.
\end{proof}

In practice, $\rho>0$ (the only connected graphs such that $\rho=0$ are
    the star-graphs), and its value tends to grow with the size of the graph.
  Therefore, Theorem 3 makes it possible to deduce from
  Empirical Result~\ref{emp:conv} the following result:
\begin{cor}[of Empirical result~\ref{emp:conv}] \label{cor-conv}
  The Markov chain converges after $O(m)$ transitions. 
\end{cor}

\begin{con} \label{con:swaps-transitions}
From now on, and in order to simplify the notations,
we will take advantage of Theorem~\ref{th:swaps-transitions} and
use the terms "edge swap" and "transition" indifferently.
\end{con}

\subsection*{Complexity}

As we have already seen, the first two steps of the random generation
(realization of the degree sequence and connection of the graph) are done
in $O(m)$ time and space.
The last step requires $O(m)$ transitions to be done (Corollary~\ref{cor-conv}).
Each transition consists in an edge swap, a simplicity test,
a connectivity test, and possibly the cancellation of the swap (\ie one more
edge swap).

Using hash tables for the adjacency lists, each edge swap and simplicity
test can be done in constant time and space. Each connectivity
test, on the contrary, needs $O(m)$ time and space.
Therefore, the $O(m)$ swaps and simplicity tests
are done in $\cs = O(m)$ time and $O(1)$ space, while
the $O(m)$ connectivity tests require $\cc = O(m^2)$ time and $O(m)$ space.
Thus, the total time complexity for the shuffle is quadratic:
\begin{equation} \label{eq:naive-comp}
  \ct_{naive} = O(m^2)
\end{equation}
while the space complexity is linear.

\medskip
One can however improve significantly this time complexity using the structures
described in \cite{henzinger,holm,thorup} to maintain
connectivity in dynamic graphs. These structures require $O(m)$
space. Each connectivity test can be performed in time
$O(\log n / \log \log \log n)$ and each simplicity test in $O(\log n)$ time.
Each edge swap then has a cost in $O(\log n (\log \log n)^3)$ time.
Thus, the space complexity is $O(m)$, and the time complexity is given by:
\begin{equation} \label{eq:dynamic-comp}
  \ct_{dynamic} = O\left(m \log n (\log \log n)^3\right)
\end{equation}

Notice however that these structures are quite intricate, and that the
constants are large for both time and space complexities.
The naive algorithm, despite the fact that it runs in $O(m^2)$ time, is
therefore generally used in practice since it has the advantage of
being extremely easy to implement.
Our contribution in this paper will be to show how it can be significantly
improved while keeping it very simple, and that it can even outperform
the dynamical algorithm.

\subsection*{Speed-up and the Gkantsidis et al. heuristics}
Gkantsidis et al. proposed a simple way to speed-up the shuffle process
\cite{gkantsidis} in the case of the naive implementation in $O(m^2)$:
instead of running a connectivity test for each transition,
they do it every $T$ transitions, for an integer $T$ called the
\emph{speed-up window}.
Thus, a transition now only consists in an edge swap and a simplicity test,
and possibly the cancellation of the swap.
If the graph obtained after these $T$ transitions is not connected anymore,
the $T$ transitions are cancelled. 

They proved that Corollary~\ref{cor:markov-conv} still holds, \ie that this
process converges to the uniform distribution, although it is
no longer composed of a single Markov chain but of a concatenation of
Markov chains~\cite{gkantsidis}.

\begin{figure}[!ht]
\begin{center}
\includegraphics[height=3cm]{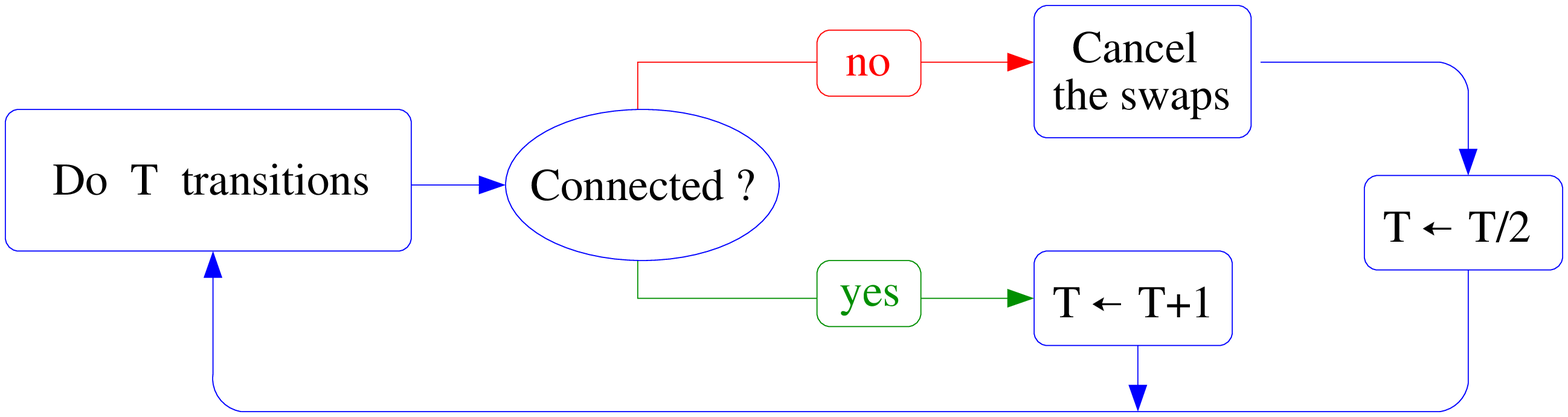}
\end{center}
\caption{Heuristics \ref{heu:gkan} (Gkantsidis et al. heuristics)}
\label{fig:gkan-heuristics}
\end{figure}

The global time complexity of connectivity tests $\cc$
is reduced by a factor $T$, but at the same time
the swaps are more likely to get cancelled:
with $T$ swaps in a row, the graph
has more chances to get disconnected than with a single one.
Let us introduce the following quantity:
\begin{df} [Success rate]\label{def:success}
  The success rate $\s(T)$ of the speed-up at a given step
  is the probability that the graph obtained after $T$ swaps is still
  connected.
\end{df}
In order to do $O(m)$ swaps, the shuffle process now requires $O(m/\s(T))$
transitions, according to Convention~\ref{con:swaps-transitions}.
The time complexity therefore becomes:
\begin{equation} \label{eq:gkan-comp}
 \ct_{Gkan} = O\left(\frac{\displaystyle m + \frac{m^2}{T}}
     {\displaystyle\s(T)}\right)
\end{equation}
The behavior of the success rate $\s(T)$ is not easily predictable.
If $T$ is too large, the graph will get
disconnected too often, and $\s(T)$ will be too small. If on the contrary
$T$ is too small, then $\s(T)$ will be large but the complexity improvement
is reduced.
To bypass this problem, Gkantsidis et al. used the
following heuristics (see Figure~\ref{fig:gkan-heuristics}).
\begin{heu}[Gkantsidis et al. heuristics]
\label{heu:gkan}
\verb|IF| \ the graph got disconnected after $T$ swaps \ 
\verb|THEN| \ $T \leftarrow T/2$ \ 
\verb|ELSE| \ $T \leftarrow T+1$
\end{heu}


Intuitively, they expect $T$ to converge to a good compromise between
a large window $T$ and a sufficient success rate $\s(T)$, depending
on the graph topology (\ie on the degree distribution).

\section{More from the Gkantsidis et al. heuristics}

The problem we address now is to estimate the efficiency of the Gkantsidis
heuristics. First, we introduce a framework to evaluate the ideal value
for the window $T$.
Then, we analyze the behavior of the Gkantsidis et al. heuristics,
and get an estimation of the difference between the speed-up
factor they obtain and the optimal speed-up factor.
We finally propose an improvement of this heuristics which reaches the
optimal. We also give experimental evidences for the obtained performance.

\subsection*{The optimal window problem}

We introduce the following quantity:
\begin{df}[Disconnection probability]
Given a graph $G$, the \emph{disconnection probability} $p$ is the probability
that the graph gets disconnected after a random edge swap.
\end{df}

\noindent
Now, let us assume the two following hypothesis:
\begin{hyp} \label{hyp:p-constant}
  The disconnection probability $p$ is constant during ~$T$ consecutive swaps
\end{hyp}
\begin{hyp} \label{hyp:no-reconnection}
  The probability that a disconnected graph gets reconnected with a random
  swap, called the \emph{reconnection probability}, is equal to zero.
\end{hyp}
\noindent
Notice that these hypothesis are not true in general. They are however
reasonable approximations in our context and will actually be confirmed
in the following. We also conduced intensive experiments which gave empirical evidence of this.
Moreover, we will give a formal explanation of the second hypothesis
in the case of scale-free graphs in Section 4.
With these two hypothesis, the success rate $\s(T)$, which is the probability
that the graph stays connected after $T$ swaps, is given by:
\begin{equation} \s(T) = (1-p)^T \end{equation}

\vspace{0.5ex}
\begin{df}[Speed-up factor]
The \emph{speed-up factor} $\r(T)$ is the {\bf expected} number
of swaps actually performed between two connectivity tests, which is $0$ if
the swaps are cancelled, and $T$ if they are not.
\end{df}
\noindent
The speed-up factor $\r(T)$, the success rate $\s(T)$ and the disconnection
probability $p$ are related as:
\begin{equation} \label{eq:factor}
  \r(T)\ =\ T \cdot \s(T)\ =\ T \cdot (1-p)^T
\end{equation}
The speed-up factor $\r(T)$ represents the \emph{actual} gain
induced by the speed-up, \ie the reduction factor of the time
complexity of the connectivity tests $\cc$.

Now, given a graph $G$ with disconnection probability $p$, the \emph{best}
window $T$ is the window that maximizes the speed-up factor $\r(T)$.
We find an optimal value $T = 1/p$, which corresponds to a success
rate $r(T) = 1/e$. Finally, we obtain the following theorem:

\begin{thm}
\label{th:ideal}
The maximal speed-up factor $\r_{max}$ is reached if and only if one of the
following equivalent conditions is satisfied:
\begin{itemize}
\item[\1] $T = \frac{1}{p}$
\item[\2] $\s(T) = e^{-1}$
\end{itemize}
The value of this maximum depends only on $p$ and is given by
  $\r_{max} = (p \cdot e)^{-1} $
\end{thm}

\subsection*{Analysis of the heuristics}

Knowing the optimality condition, we tried to estimate the performance of the
Gkantsidis et al. heuristics. Considering $p$ as given, the evolution of the
window $T$ under these heuristics leads to :
\begin{thm}
 \label{th:gkan}
 The speed-up factor $\r_{Gkan}(p)$ obtained with the Gkantsidis heuristics
 verifies:
 $$
  \forall \epsilon>0, \qquad
    \r_{Gkan} = o\left((\r_{max})^{\frac{1}{2}+\epsilon}\right)
  \quad when \quad p \to 0
 $$
\end{thm}

\begin{proof}[Sketch of proof]
We give here a simple
mean field approximation leading to the stronger, but approximate
result: $\r_{Gkan} = \Omega\left(\sqrt{\r_{max}}\right)$.
The proof of Theorem~\ref{th:gkan}, not detailled here,
follows the same idea.

\medskip
\noindent
Given the window $T_t$ at step $t$, we obtain an expectation for $T_{t+1}$
depending on the succes rate $\s(T_t)$:
\[ \overline{T_{t+1}} = \s(T_t) (T_t+1) + (1-\s(T_t))\frac{T_t}{2} \]
We now suppose that $T$ eventually reaches a mean value $\textbf{T}$. We then
obtain:
\[ \mt = (1-p)^{\mt} (\mt+1) + \left(1-(1-p)^{\mt}\right) \frac{\mt}{2} \]
which leads to
\[ \frac{\mt}{2} = \frac{1}{1-(1-p)^{\mt}} - 1\]
Therefore, if $p \to 0$ we must have $\mt \to \infty$, thus
$1-(1-p)^{\mt} \to 0$ and finally $1-(1-p)^{\mt} \sim p \mt$. Therefore:
\[ \mt \sim \sqrt{\frac{2}{p}} \]
which gives, with Equation~\ref{eq:factor} and Theorem~\ref{th:ideal},
  still for $p \to 0$:
\begin{equation} \label{eq:gkan}
  \r_{Gkan} \sim \sqrt{2 e \cdot\r_{max}}
\end{equation}
   
\end{proof}

Intuitively, this Theorem means that the Gkantsidis et al. heuristics
is too pessimistic: when the graph gets disconnected, the decrease of $T$
is too strong; conversely, when the graph stays connected, $T$ grows too slowly.
By doing so, one obtains a very high success rate (very close to 1), which
is not the optimal (see Theorem~\ref{th:ideal}).

\subsection*{An optimal dynamics}

To improve the Gkantsidis et al. heuristics we propose the following
  one (with two parameters $q^-$ and $q^+$)~:
\begin{heu} \label{heu:fab}
\verb|IF| \ the graph got disconnected after $T$ swaps \ 
\verb|THEN| \ $T \leftarrow T \cdot (1-q^-)$ \ 
\verb|ELSE| \ $T \leftarrow T \cdot (1+q^+)$
\end{heu}

The main idea was to avoid the linear increase in $T$, which is too slow,
and to allow more flexibility between the two factors $1-q^-$ and $1+q^+$.

\begin{thm}
With this heuristics, a constant $p$, and for $q^+ , q^-$ close enough to $0$,
the window $T$
converges to the optimal value and stays arbitrarily close to it with
arbitrarily high probability if and only if
\begin{equation} \label{eq:idealcondition}
  \frac{q^+}{q^-} = e-1
\end{equation}
\end{thm}

\begin{proof}[Sketch of proof]
If the window $T$ is \emph{too large}, the success rate
$\s(T)$ will be \emph{small}, and $T$ will decrease.
Conversely, a \emph{too small} window $T$ will grow.
This, provided that the factors $1+q^+$ and $1-q^-$ are close enough to $1$,
ensures the convergence of $T$ to a mean value $\textbf{T}$.
Like in the proof of Theorem~\ref{th:gkan}, we have:
$$ \mt = \s(\mt)(1+q^+)\mt + (1-\s(\mt))(1-q^-)\mt $$
This time, the error made by this approximation can be as small as one wants
by taking $q^+$ and $q^-$ small enough, so that $T$ stays close to its
mean value $\mt$. It follows that:
$$ \s(\mt) = \frac{q^-}{q^+ + q^-} $$
This quantity is equal to $e^{-1}$ (optimality condition \2 of
 Theorem~\ref{th:ideal}) if and only if $\frac{q^+}{q^-} = e-1$
\end{proof}

\subsection*{Experimental evaluation of the new heuristics}

To evaluate the relevance of these results, based on
Hypothesis~\ref{hyp:p-constant} and \ref{hyp:no-reconnection},
and dependent on a constant value of $p$ (which is not the
    case, since the graph continuously changes during the shuffle)
we will now compare empirically the three following heuristics:
\begin{enumerate}
\item The Gkantsidis et al. heuristics
  (Fig.~\ref{fig:gkan-heuristics}, Heuristics~\ref{heu:gkan})
\item Our new heuristics (Heuristics~\ref{heu:fab})
\item The \emph{optimal} heuristics: at every step, we compute the window $T$
  giving the maximal speed-up factor $\r_{max}$.\footnote{Note that the
  heavy cost of this operation prohibits its use as a heuristics, out of this
  context. It only serves as a reference.}
\end{enumerate}
We compared the average speed-up factors obtained with these three heuristics
(respectively $\r_{Gkan}$, $\r_{new}$ and $\r_{max}$) for the generation of
graphs with various heavy tailed\footnote{
  To obtain heavy tailed distributions, we used power-law like
  distributions: $P(X=k) = (k+\mu)^{-\alpha}$,
  where $\alpha$ represents the ``heavy tail'' behavior,
  while $\mu$ can be tuned to obtain the desired average $z$.
}
degree sequences.
We used a wide set of parameters, and all the results were consistent
with our analysis: the average speed-up factor $\r_{Gkan}$ obtained with the
Gkantsidis et al. heuristics
behaved asymptotycally like the square root of the optimal, and our average
speed-up factor $\r_{new}$ always reached at least $90\%$ of the optimal $\r_{max}$.
Some typical results on heavy-tailed distributions with $\alpha = 2.5$
and $\alpha = 3$ are shown below.

\begin{table}[!ht]
\label{tab:speedupfactors}
\begin{center}
\begin{tabular}{|c|c|c|c|}
  \multicolumn{4}{c}{$\alpha=2.5$} \\
  \hline
  $z$ &$\r_{Gkan}$&$\r_{new}$ & $\r_{max}$\\
  \hline
  2.1 & 0.79 & 0.88  & 0.90  \\
  3   & 3.00 & 5.00  & 5.19  \\
  6   & 20.9 & 112   & 117   \\
  12  & 341  & 35800 & 37000 \\
  \hline
\end{tabular}
\hspace{1cm}
\begin{tabular}{|c|c|c|c|}
  \multicolumn{4}{c}{$\alpha=3$} \\
  \hline
  $z$ &$\r_{Gkan}$&$\r_{new}$ & $\r_{max}$\\
  \hline
  2.1 & 1.03 & 1.20  & 1.26  \\
  3   & 5.94 & 12.3  & 12.4  \\
  6   & 32.1 & 216   & 234   \\
  12  & 578  & 89800 & 91000 \\
  \hline
\end{tabular}
\ccaption{Average speed-up factors for various values of the average degree $z$.
We limited ourselves to $n=10^4$ because the computations are quite expensive,
in particular concerning $\r_{Gkan}$ and $\r_{max}$.}
\end{center}
\end{table}

These experiments show that our new heuristics is very close to the optimal.
Thus, despite the fact that $p$ actually varies during the shuffle, our
heuristics react fast enough (in regard to the variations of $p$) to get
a good, if not optimal, window $T$. 
We therefore obtain a success rate $\s(T)$ in a close range around $e^{-1}$.
From Equation~\ref{eq:gkan-comp} and Theorem~\ref{th:ideal}, we obtain
the following complexity for the shuffle:
\begin{equation} \label{eq:opt-complexity}
  \ct_{new} = O\left( m + <p> \cdot m^2\right)
\end{equation}
(where $<p>$ is the average value of $p$ during the shuffle),
instead of the $O\left(m + \sqrt{<p>} \cdot m^2 \right)$ complexity of
the Gkantsidis et al. heuristics, also obtained from
Eq.~\ref{eq:gkan-comp} and Th.~\ref{th:ideal}.
Further empirical comparisons of the two heuristics will be provided in
the next section, see Table~\ref{tab:times}.

Our complexity $C_{new}$, despite the fact that it is asymptotically
still outperformed by the complexity of the dynamic connectivity algorithm
$C_{dynamic}$ (see Eq. \ref{eq:dynamic-comp}),
may be smaller in practice if $p$ is small enough.
For many graph topologies corresponding to real-world networks, especially
graphs having a quite high density (social relations, word co-occurences, WWW),
and therefore a low disconnection probability, our algorithm represents an alternative that may behave faster, and which implementation is much easier.

\section{A log-linear algorithm ?}
We will now show that, in the particular case of heavy-tailed degree
distributions like the ones met in practice \cite{faloutsos, newman-review},
one may 
reduce the disconnection probability $p$ at logarithmic cost, thus reducing
dramatically the complexity of the connectivity tests.
We first outline the main idea, then we present empirical tests
showing the asymptotical behavior of the disconnection probability:
this leads us to a conjecture, strongly supported by both intuition,
experiments and formal arguments, from which we obtain a $O(n \log n)$
algorithm.
We finally improve this algorithm, which makes us expect a $O(n \log \log n)$
complexity.

\subsection*{Guiding principle}
In a graph with a heavy-tailed degree distribution, most vertices 
have a very low degree. This means in particular that,
  swapping two random edges, one has
a significant probability to connect two vertices of degree 1 together,
 creating an isolated component of size 2.
One may also create small components of size 3, and so on.
Conversely, the non-negligible number of vertex of high degree
form a robust core, so that it is very unlikely that a random swap
creates two large disjoint components.
Therefore, an heavy-tailed distribution implies that when a swap
disconnects the graph, it creates in general a \emph{small} isolated
component rather than a large one.
\begin{df}[Isolation test] \label{def:isolation}
  An isolation test of width $K$ on vertex $v$ tests wether this vertex
  belongs to a connected component of size lower than or equal to $K$.
\end{df}
To avoid the disconnection, we now do an isolation test for every
transition, just after the simplicity test.
If this isolation test returns \verb|true|, we cancel the swap rightaway.
This way, we detect at low cost (as long as $K$ is small)
  a significant part of the disconnections.
As before, we will use Convention~\ref{con:swaps-transitions}, considering
that \emph{every} transition corresponds to a valid swap, \ie a swap that
passes both the simplicity and isolation tests.

The disconnection probability $p$ is now the probability that after $T$
swaps which passed the isolation test, the graph gets disconnected.
It is straightforward to see that $p$ is decreasing with $K$,
even if the relation between them is yet to establish.
Therefore, given a graph,
 the success rate $\s$ now depends on both $T$ and $K$.

\subsection*{Empirical study of $p$}
\begin{figure}[!hb]
\begin{center}
 \includegraphics[height=4cm]{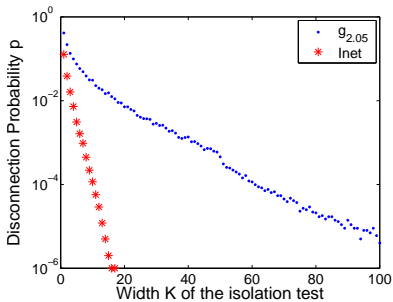}
 \end{center} 
 \caption{Empirical behavior of $p(K)$ for two degree distributions}
\label{fig:K}
\end{figure}

Since the disconnection probability $p$ now depends on $K$, and in order to
study this relation, we will denote it by $p(K)$ in this subsection.
We ran extensive experiments on a very large variety of heavy-tailed degree
distributions and graph sizes, as well as real-world network degree
distributions.
Results are shown in Figure~\ref{fig:K} for the degree distribution of the
Internet backbone topology presented in \cite{govindan} (Inet) and for the
heavy-tailed degree distribution with the values for $z$ and $\alpha$ that
gave the worst results ($g_{2.05}$),
\ie the largest $p(K)$. This worst-case distribution had average degree
$z=2.05$ and exponent $\alpha=2.1$.

We finally state the following conjecture and assume it is true in the sequel:
\begin{conj} \label{conj:K-exp}
  The average disconnection probability
  for random simple connected graphs with heavy-tailed degree distributions
  decreases {\bf exponentially} with $K$: $p(K) = O(e^{-\lambda K})$ for
  some positive constant $\lambda$ depending on the distribution,
  and not on the size of the graph.
\end{conj}

\subsection*{The final algorithm}
Let us introduce the following quantity:
\begin{df}[Characteristic isolation width] \label{df:characterisitc-K}
  The characteristic isolation width $K_G$ of a graph $G$ having $m$ edges
  is the minimal \emph{isolation test width} $K$ such that the disconnection
  probability $p(K)$ verifies $p(K) < 1/m$.
\end{df}

\noindent
This leads naturally to:
\begin{lem} \label{lm:apply-K}
  Applying the shuffle process to a graph $G$ having at least $10$ edges,
  with an isolation test width
  $K \geq K_G$, and a period $T$ equal to $m$, we obtain a success rate
  $\s$ larger than $\frac{1}{3}$.
\end{lem}

\begin{proof}
  Even without Hypothesis~\ref{hyp:no-reconnection}, the success rate is
  always \emph{greater than or equal} to $(1-p)^T$. Choosing $K \geq K_G$ and
  $T = m$, we obtain:
  $$ \s \geq \left(1-\frac{1}{m}\right)^m $$
  which is larger than $\frac{1}{3}$ for $m \geq 10$
\end{proof}

\noindent
Moreover, still assuming Conjecture~\ref{conj:K-exp}, and
because $m=O(n)$ for the degree distributions we consider:
\begin{lem} \label{lm:K-log}
  For a given degree distribution, the characteristic isolation width $K_G$
  of random graphs of size $n$ is in $O(\log n)$
\end{lem}
  
\noindent
It follows that:
\begin{thm} \label{log-linear}
  For a given degree distribution, the shuffle process
  for graphs of size $n$ has complexity $O(n \log n)$ time and $O(m)$ space.
\end{thm}

\begin{proof}
  Let us define the procedure \verb|shuffle(G)| as follows:
  \begin{enumerate}
  \item set $K$ to $1$,
  \item save the graph $G$,
  \item do $m$ edge swaps on $G$ with isolation tests of width $K$,
  \item if the obtained graph is connected, then return it,
  \item else, restore $G$ to its saved value, set $K$ to $2 \cdot K$,
    and go back to step 2.
  \end{enumerate}
  This procedure returns a connected graph obtained after applying $m$
  edge swaps to $G$.
  Lemma~\ref{lm:apply-K} and~\ref{lm:K-log} ensure that this procedure 
  ends after $O(\log \log n)$ iterations with high probability.
  Moreover, the cost of iteration $i$ is $O(2^i \cdot m)$, since
 dominating complexity comes from the $O(m)$ isolation tests of width $2^i$.
  Therefore, we obtain a global complexity of $O(m \log n)$ time.
  We have $m = O(n)$, so that the
  complexity is finally $O(n \log n)$ time. The space complexity is
  straightforward.
\end{proof}

\begin{figure}[!ht]
\begin{center}
\includegraphics[height=5cm]{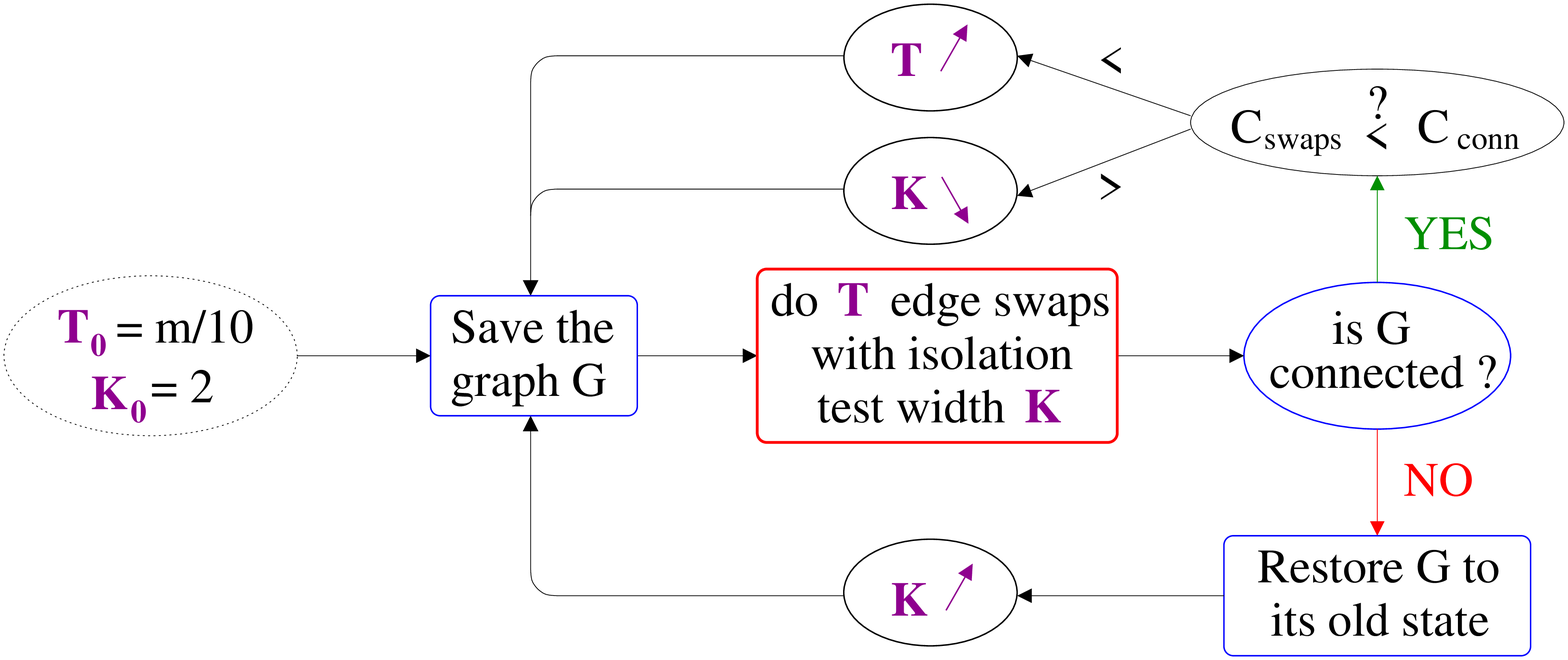}
\end{center}
\ccaption{Our final heuristics used to adjust the isolation test width $K$
  and the
  window $T$ in an implementation of the log-linear algorithm.}
\label{fig:KT}
\end{figure}

One can easily check that the heuristics presented in 
Figure~\ref{fig:KT} is at least as efficient as the one presented in this proof.
It aims at 
equilibrating $\cs$ and $\cc$ by dynamically adjusting the isolation test width
$K$ and the window $T$, keeping a high success rate $\s(K,T)$ and a 
large window $T$ (here we impose $T \geq \frac{m}{10}$).

We compare in Table~\ref{tab:times} typical running times obtained for various
sizes and for an heavy-tailed degree distribution with the naive algorithm,
the Gkantsidis et al. heuristics, our improved version of this heuristics,
  and our final algorithm. 
  Notice that our final algorithm allows to generate massive graphs
  in a very reasonable time, while the previously used heuristics
  could need several weeks to do so.
  The limitation for the generation of massive graphs
  now comes from the memory needed to store the graph more than the
  computation time.
  Implementations are provided at 
  \href{http://www.liafa.jussieu.fr/~fabien/generation}{\cite{www-generation}}.

\begin{table}[!ht]
\begin{center}
\begin{tabular}{|c|r|r|r|r|}
  \hline
  $m$ & Naive & Gkan. heur. & Heuristics 2 & Final algo. \\
  \hline
  $10^3$ & 0.51s & 0.02s & 0.02s & 0.02s \\
  $10^4$ & 26.9s & 1.15s & 0.47s & 0.08s \\
  $10^5$ & 3200s & 142s  &   48s &  1.1s \\
  $10^6$ & \~ 4$\cdot 10^5$s & \~ 3$\cdot 10^4$s & 10600s & 25.9s \\
  $10^7$ & \~ 4$\cdot 10^7$s & \~ 3$\cdot 10^6$s & \~ $10^6$s & 420s \\
  \hline
\end{tabular}
\end{center}
\ccaption{Average computation time for the generation of graphs of
  various sizes with an heavy-tailed degree distribution of
  exponent $\alpha = 2.5$ and average degree $z = 6.7$
  (on a Intel Centrino 1.5MHz with 512MB RAM).}
\label{tab:times}
\end{table}

\subsection*{Towards a \protect{$O(n \log \log n)$} algorithm ?}
The isolation tests are typically breadth- or depth-first searches that stop when they
have visited $K+1$ vertices. Taking advantage of the heavy-tailed
degree distribution, we may be able
to reduce their complexity as follows: if the search meets a vertex of degree greater than $K$, it can stop because it means that the component
is large enough. This is a first improvement.

Moreover, if the search is processed in an appropriate way,
like a depth-first search directed at the neighbour of highest degree,
it may reach a vertex of degree greater than $K$ in only a few steps.
Several recent results indicate that searching a vertex of degree at least
$K$ in an heavy-tailed network takes $O(\log K)$ steps in
average~\cite{sarshar,adamic-short,adamic}.
Thus, running an isolation test after an edge swap that
\emph{did not disconnect}
the graph would be done in $O(\log \log n)$ time instead of $O(\log n)$.

In the case of a swap that disconnected the graph, we also have the
following result:
\begin{lem}
  For a given heavy-tailed degree distribution,
  the expected complexity of an isolation test, knowing that it returned
 \verb|true|, is constant.
\end{lem}
\begin{proof}
  Knowing that the test returned \verb|true|, the probability $s_i$
  that the isolated component had a size equal to $i$ is given by:
  $$ s_i = \frac{p(i)-p(i+1)}{p(0)-p(K+1)}$$
  where $p(k)$ is the disconnection probability for an isolation test width $k$.
  Using Conjecture~\ref{conj:K-exp} that assumes an exponential decrease
  of $p(k)$, it follows that
  the expectation of the size of this isolated component is $O(1)$.
\end{proof}

Finally, the complexity of \emph{any} isolation test would be $O(\log \log n)$ time,
  so that the global complexity would become $O(n \log \log n)$ time.

\section{Conclusion}
Focusing on the speed-up method introduced by Gkantsidis et al. for the Markov
chain Monte Carlo algorithm,
we introduced a formal background allowing us to show that this
heuristics is not optimal in its own family. 
We improved it in order to reach the optimal, and empirically
confirmed the results.

Going further, we then took advantage
of the characteristics of real-world networks to introduce an original
method allowing the generation of random simple connected graphs 
with heavy-tailed degree distributions in $O(n \log n)$ time
and $O(m)$ space. It outperforms the previous best known methods, and
has the advantage of being extremly easy to implement.
We also have pointed directions for further 
enhancements to reach a complexity of $O(n \log \log n)$ time.
The empirical measurement of the performances of our methods show that it yields significant progress.
We provide an implementation of this last algorithm
\href{http://www.liafa.jussieu.fr/~fabien/generation}{\cite{www-generation}}.

Notice however that the last results rely on a conjecture, for which we gave several arguments and strong empirical evidences, but were unable to prove.

\bibliographystyle{plain}
\bibliography{biblio}

\newpage
\appendix
 
\section{Evaluation of the bias of the common method}
The ``common method'' to generate random simple connected graphs with
a prescribed degree sequence is the following :
\begin{enumerate}
\item Generate a graph $G$ with the Molloy and Reed model~\cite{molloy-reed}. 
\item Remove the multiple edges and loops, obtaining a simple graph $G_S$.
\item Keep only the largest connected component, obtaining a simple connected 
subgraph $G_{CS}$ 
\end{enumerate} 
In the following, we also call $G_C$ the subgraph obtained by step 3 without
step 2 ($G_C$ is the non-simple giant connected component of $G$). It is clear
that $G_S$, $G_C$ and $G_{CS}$ are different from $G$. We provide here
experimental evidences that this difference is significant.
Since our model doesn't suffer of any such bias, as it is simple and connected
from the beginning, we recommend its use for
anyone who needs to generate random simple connected graphs with a
prescribed degree sequence.

\subsection*{Notations}
We call $N$ the number of vertices in $G$,
$M$ the number of edges and $Z$ the average degree. Likewise,
$N_C$, $N_S$, $N_{CS}$,
$M_C$, $M_S$, $M_{CS}$,
$Z_C$, $Z_S$ and $Z_{CS}$
refer respectively to $G_C$, $G_S$ and $G_{CS}$.

\begin{figure}[!hb]
\begin{center}
\includegraphics[width=0.25\textwidth]{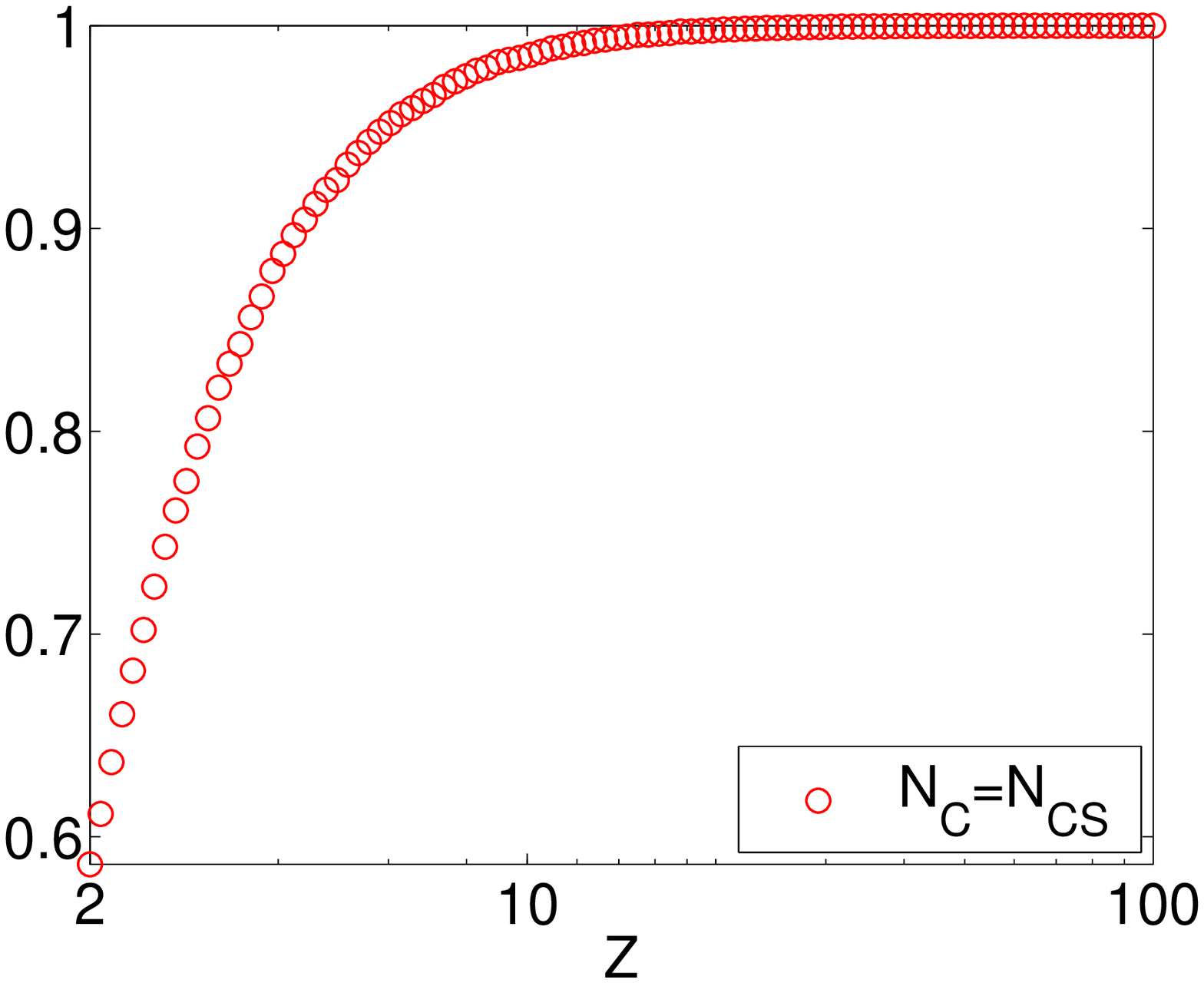}
\hspace{0.5cm} 
\includegraphics[width=0.25\textwidth]{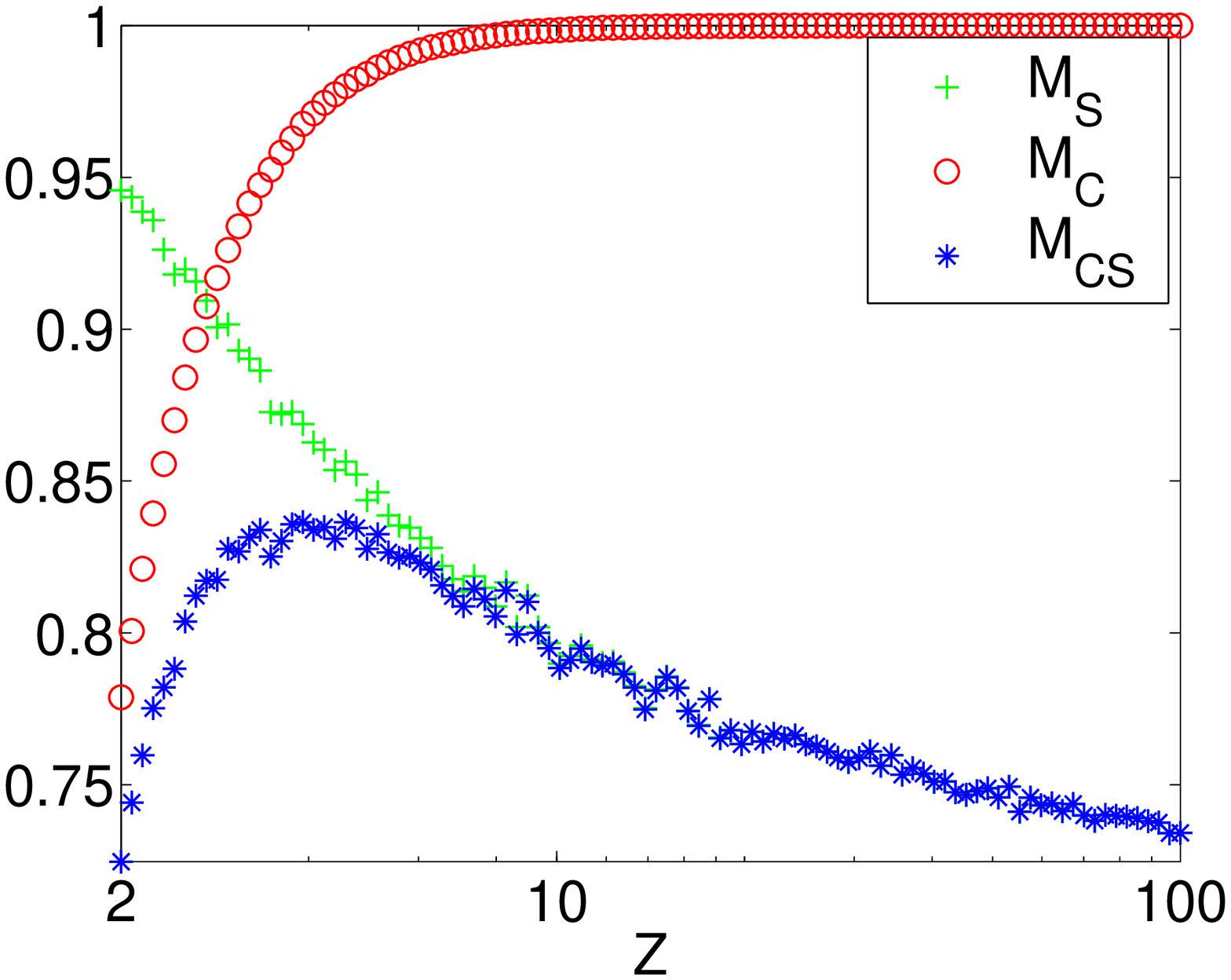}
\hspace{0.5cm} 
\includegraphics[width=0.25\textwidth]{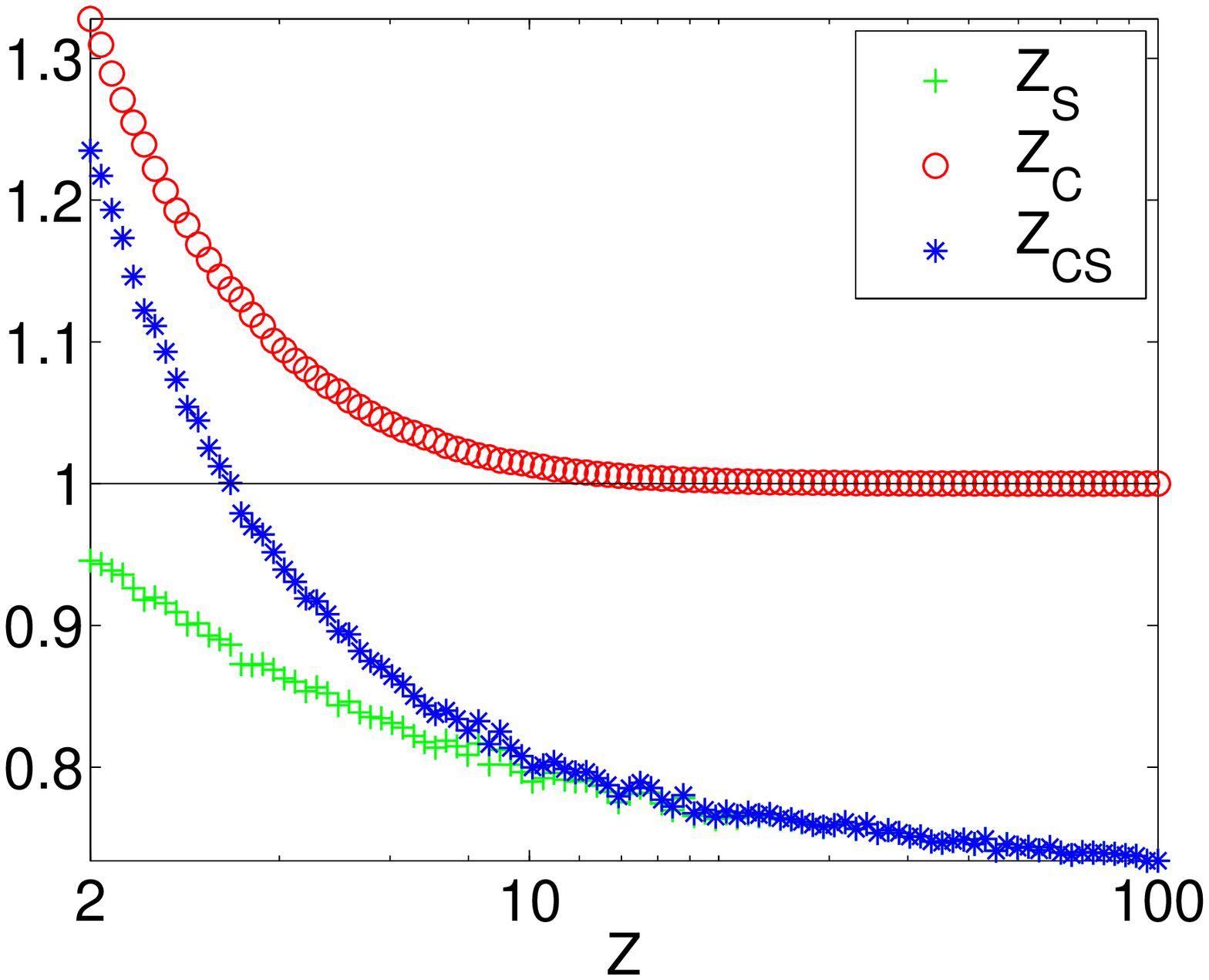}

\includegraphics[width=0.25\textwidth]{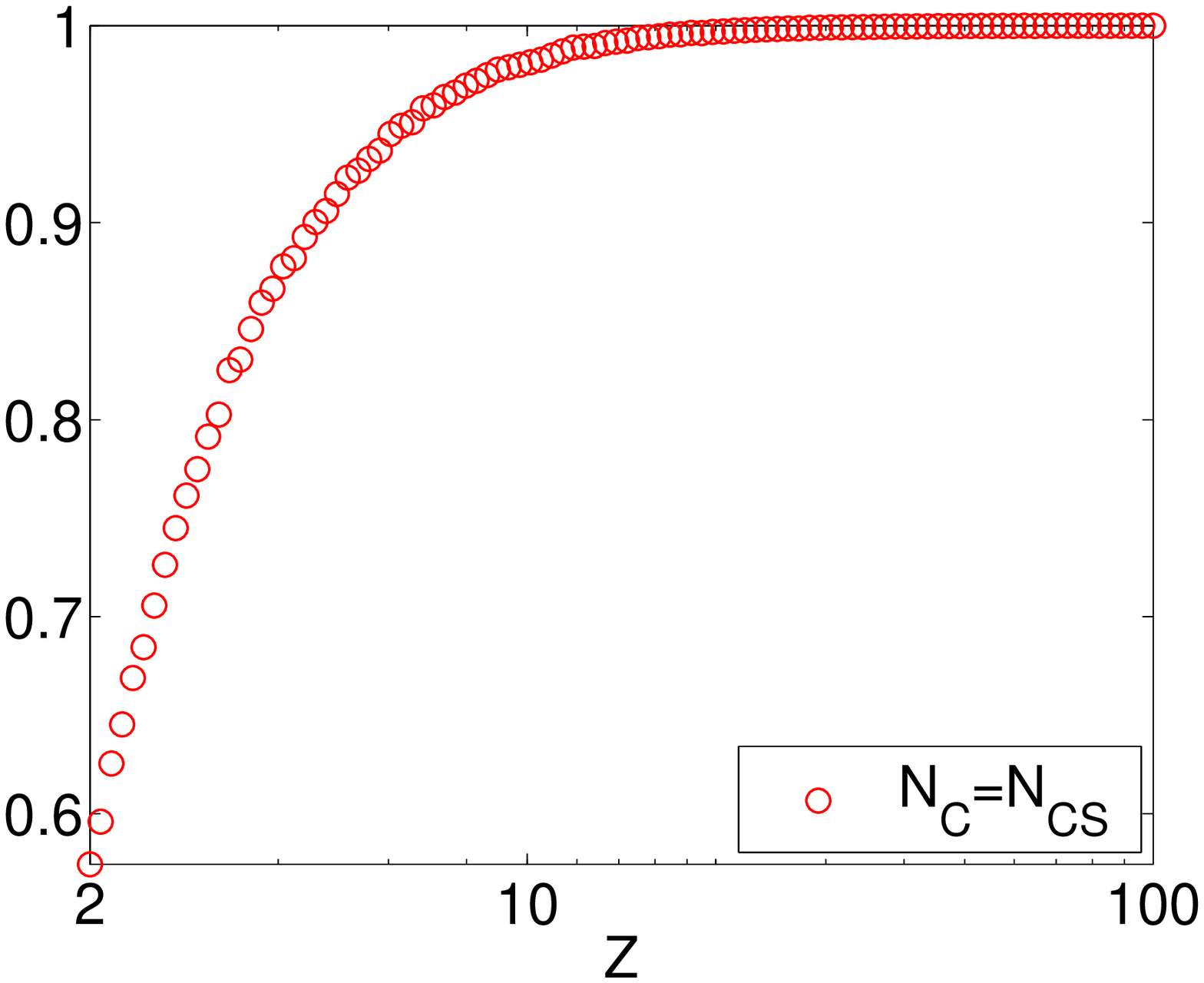}
\hspace{0.5cm} 
\includegraphics[width=0.25\textwidth]{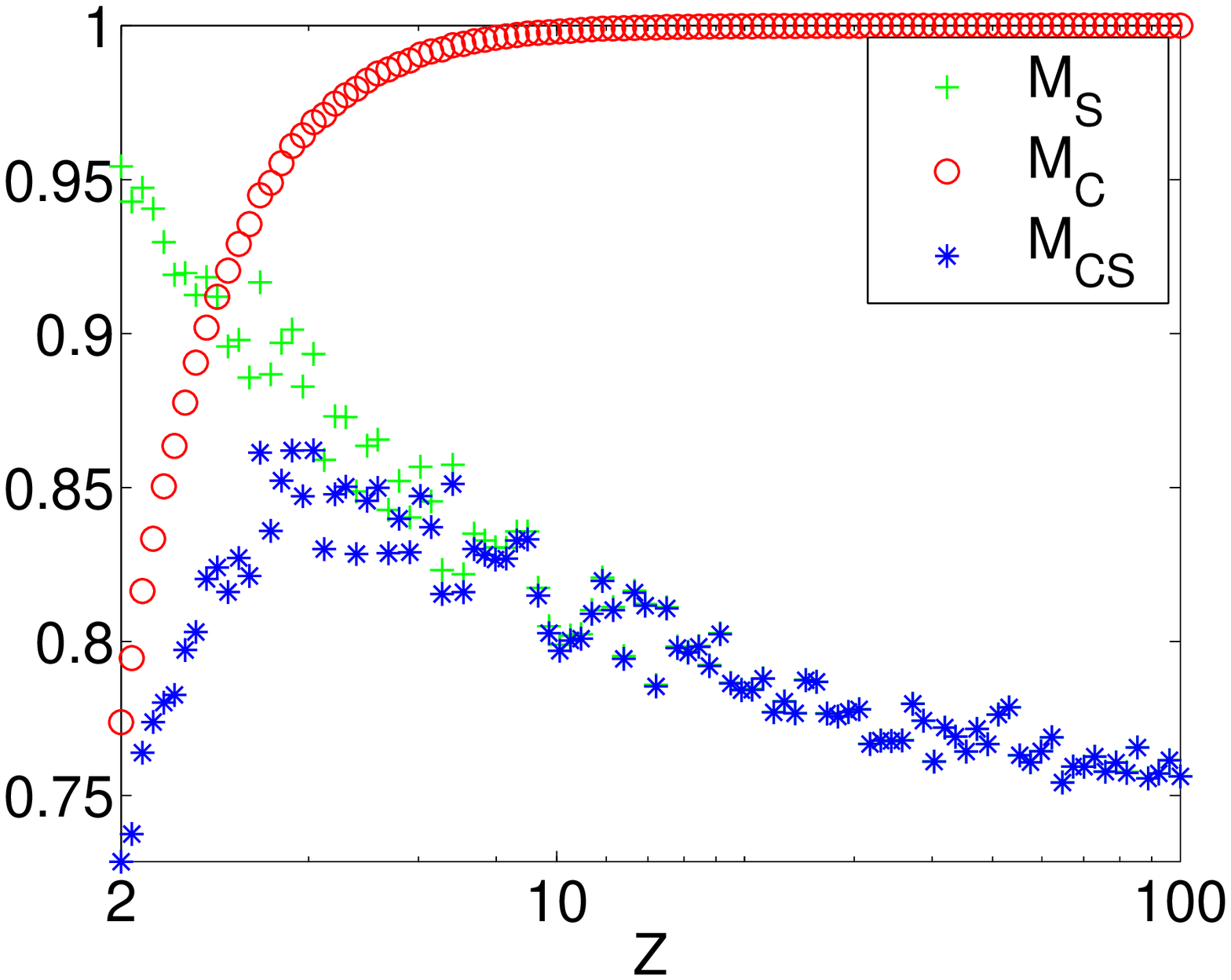}
\hspace{0.5cm} 
\includegraphics[width=0.25\textwidth]{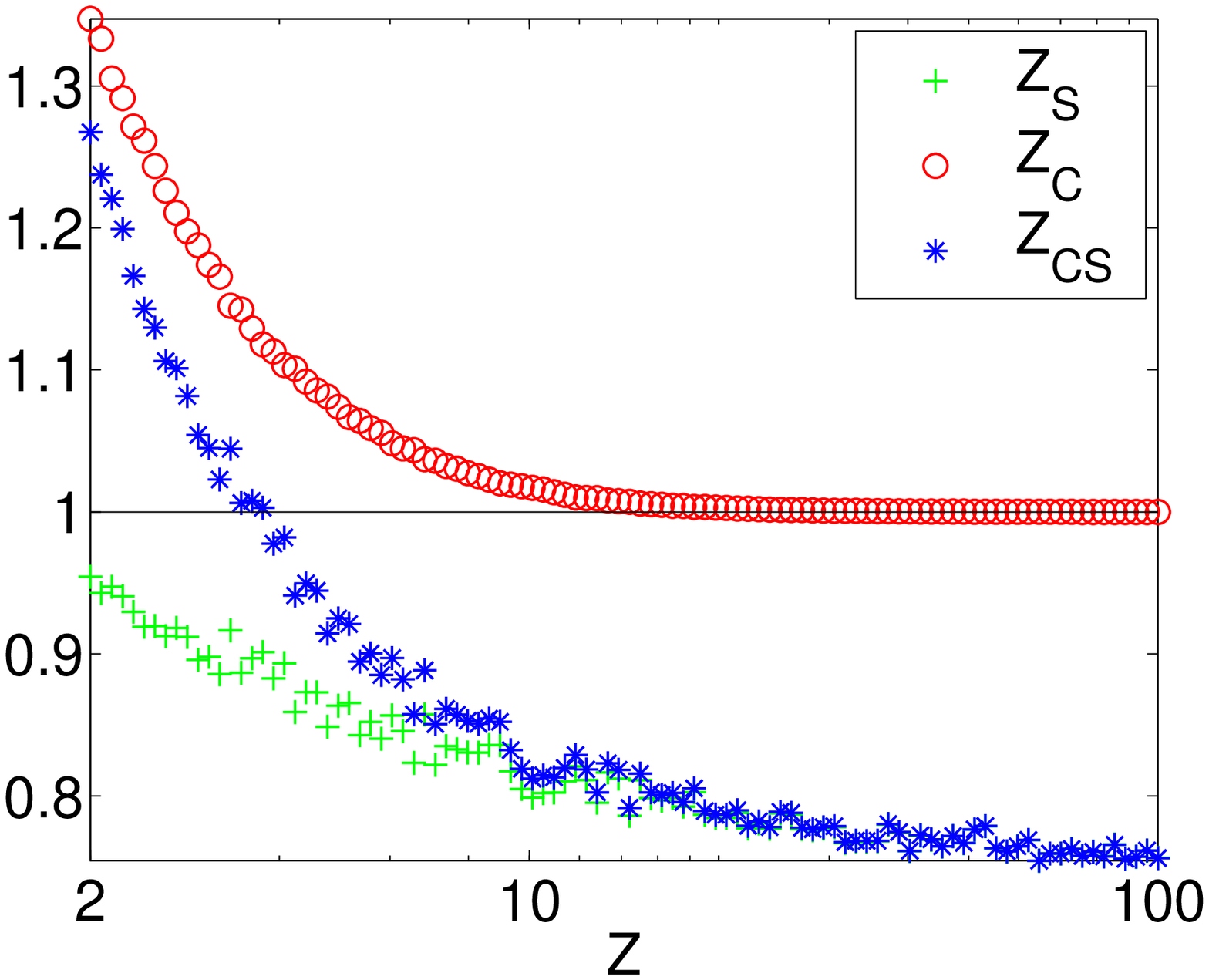}

\includegraphics[width=0.25\textwidth]{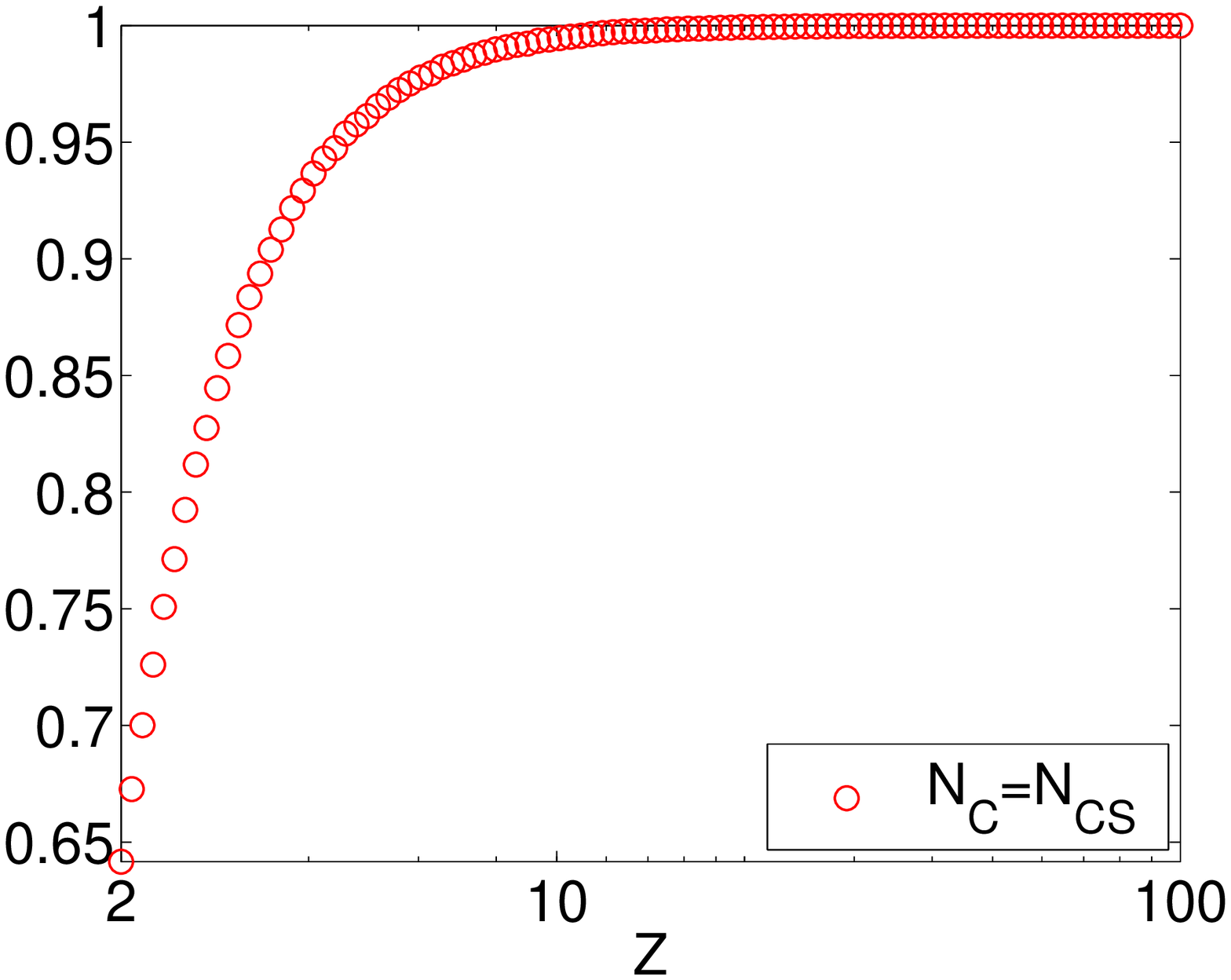}
\hspace{0.5cm} 
\includegraphics[width=0.25\textwidth]{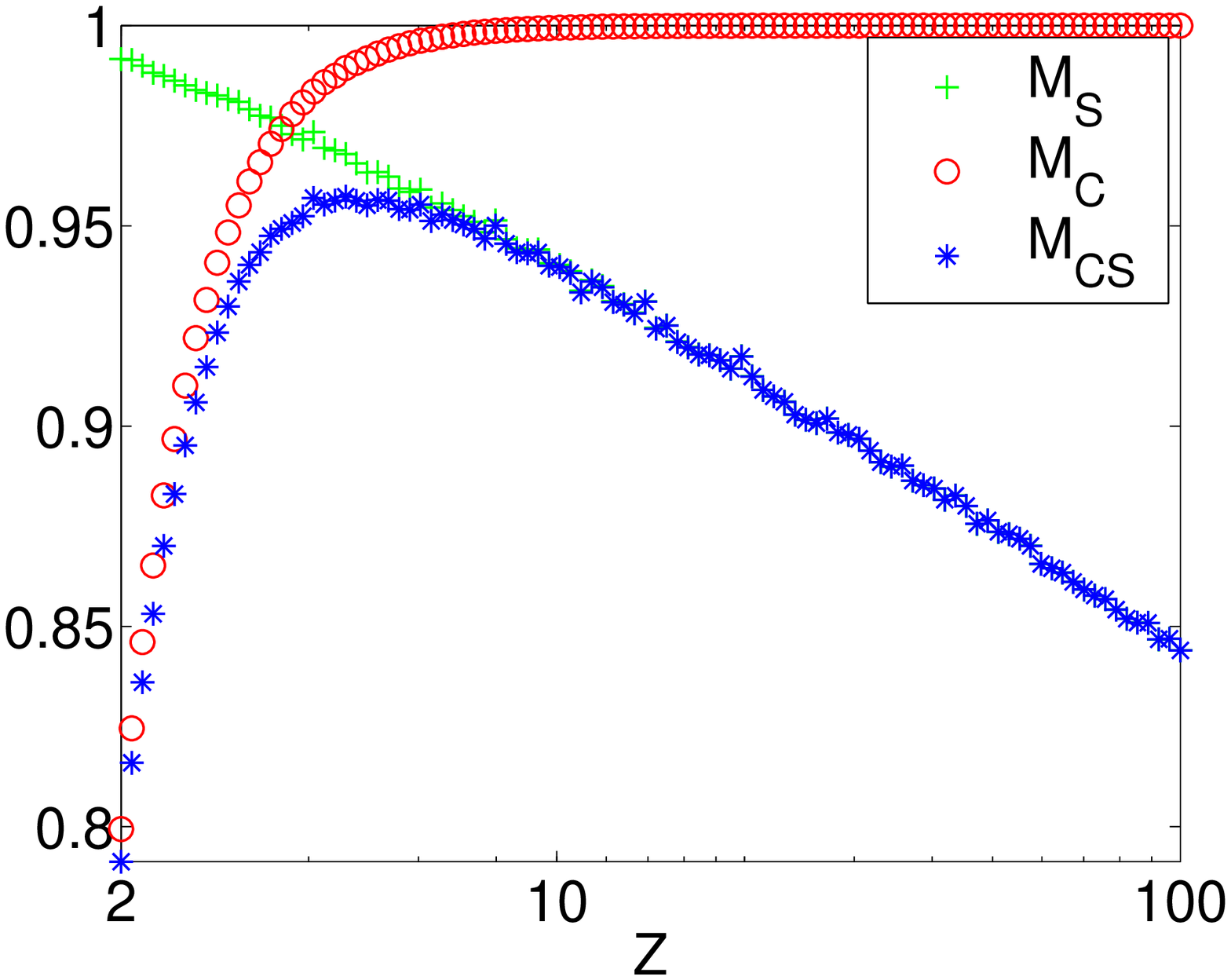}
\hspace{0.5cm} 
\includegraphics[width=0.25\textwidth]{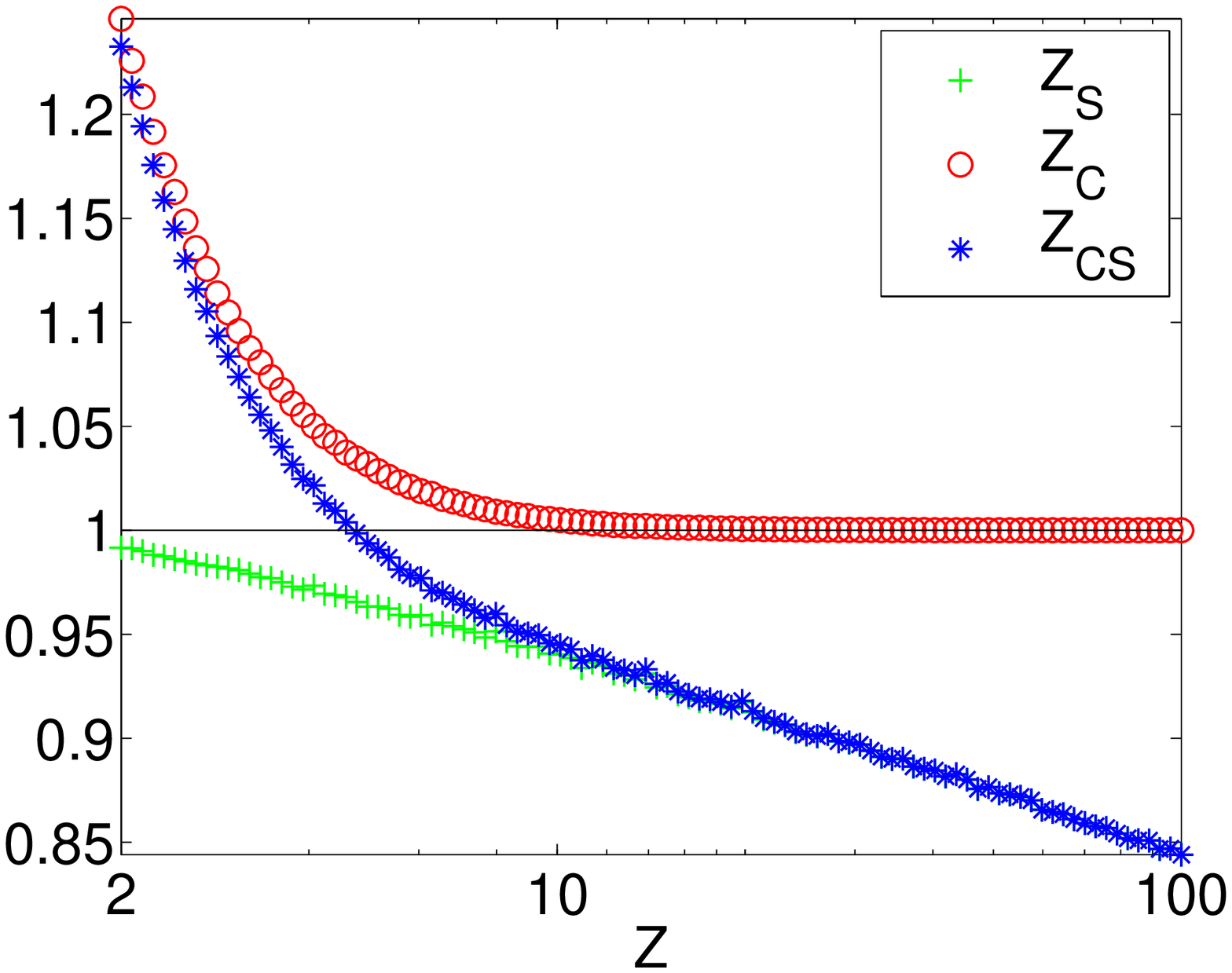}
\end{center}
\ccaption{
Comparison of the number of vertices (left), the number of edges (center)
and the average degree (right) in the graphs $G_S$, $G_C$ and $G_{CS}$,
for various values of the average degree $Z$ of the original graph $G$.
We used heavy-tailed degree distributions with
$N=10^4,\alpha=2.1$ (top), 
$N=10^5,\alpha=2.1$ (middle) 
and $N=10^4,\alpha=2.5$ (bottom).
}
\end{figure}

\subsection*{Plots}
To quantify the modifications caused by the removal of multiple edges and/or
the restriction to the giant connected component, we plotted the number
of vertices, the number of edges and the average degrees of the concerned
subgraphs $G_C$, $G_S$ and $G_{CS}$ against the average degree of $G$.
In each plot, the three curves refer to $G_C$ (red circles), $G_S$ (green plus) and $G_{CS}$(blue stars).
The quantities are {\bf normalized} so that a value of $1$ represents the
value of the concerned quantity in $G$.

Notice that, since $N_S$ is always equal to $N$ (the removal of edges
by itself does not change the number of vertices), we only plotted $N_C$, which
is also equal to $N_{CS}$.

\subsection*{Discussion}
Many things can be observed from those plots. In particular :
\begin{itemize}
\item
The left and middle plots show clearly that one loses a significant part of the
graph when performing multiple edge removal, restricting to the giant
component, or both.

\item
The similarity between the plots at the top and in the middle show that
the size $N$ has very little, if any, influence on this loss.
The only noticeable
difference comes fom the fact that the top plots, due to their lower
computation costs, were averaged on more instances than the middle ones.

\item
The bottom plots are closer to $1$, meaning that the bias is less significant.
This is due to the greater exponent $\alpha$, causing the heavy-tailed
degree distribution to be less heterogenous. Thus, less vertices have very low
degree (these ones get more likely removed in $G_C$) or very high degree
(these ones are more likely to get many edges removed in $G_S$).

\item
The left part of the plots (low average degree $Z$) show a significant loss
of vertices in $G_C$. This is of course because the more edges we have,
the bigger the giant connected component is.
On the other hand, the right part of the plots (high average degree $Z$) show
an increasing loss of edges due to the removal of more multiple edges.

\item
The plots on the right-hand side show that two opposite biases act on the
average degree $Z_{CS}$ of $G_{CS}$: the multiple edges removals tends to lower
it, while the removal of vertices that don't belong to the giant component tends
to raise it (since these vertices more likely have a low degree).

\end{itemize}

\subsection*{Conclusions}
We showed that the bias caused by the two last steps of the ``common method''
is significant, not only on the size of the graph but also on its properties,
like the average degree. These biases should therefore cause the deviation
of many other properties. Our model, which respect exactly the degree
sequence given at the beginning, represents a reference that may be used
to better quantify these deviations. Its simplicity and efficiency should
also convince users to implement it (or to use our implementation, available at
\href{http://www.liafa.jussieu.fr/~fabien/generation}{\cite{www-generation}}).
Notably, it provides an easy way to separate the properties of the known
models, like the Barab\`asi-Albert one, in two groups: the ones that come
from the degree distribution only, and the ones that come from the model itself.

\end{document}